\documentclass[aps,pra,reprint,floatfix]{revtex4-2}
\usepackage[colorlinks,linkcolor=blue,citecolor=blue]{hyperref}
\usepackage{amsmath,bm}
\usepackage{balance} 
\usepackage{graphicx}
\usepackage[T1]{fontenc}

\newcommand{\na}{\bm\nabla}
\renewcommand{\Im}{{\rm Im\,}} 
\renewcommand{\Re}{{\rm Re\,}}

\begin{document}
\title{Polarization switching of double azimuthal backflow in two transverse orbital angular momentum superpositions}
\author{Zhen-Lai Wang}
\email{wangzhenlai@hbpu.edu.cn}
\affiliation{Center for Fundamental Physics and School of Artificial Intelligence, Hubei Polytechnic University, Huangshi 435003, China}
\date{\today}

\begin{abstract}
Optical backflow in the azimuthal direction can be generated by the interference of two rotating-wave structured light beams with distinct transverse topological charges and equal intensity. Here we show that the interference reverses both azimuthal components of the Poynting and the canonical momenta. Although these two momenta are not generally tied, they reverse jointly here: both acquire azimuthal backflow in a common angular sector, while their radial intervals can differ. The double backflow is switched solely by the polarizations of the two beams, without amplitude tuning. A single polarization overlap enters the interference terms of both momenta but not the single-beam terms of either, so that matched polarizations switch the double backflow on whereas mismatched ones switch it off. The double backflow is periodic in the azimuthal direction, with its angular period locked to the topological charge difference and independent of the individual charges. Our results establish a vectorial mechanism for polarization-gated azimuthal backflow, with potential applications in polarization-switchable optical manipulation.
\end{abstract}

\maketitle

\section{Introduction}

Backflow is a counterintuitive wave-interference phenomenon in which a wave constructed from forward-going components can carry a local flux in the opposite direction. It was first systematically identified in quantum mechanics, where a wave packet consisting of purely positive momenta exhibits a nonzero probability current in the opposite direction~\cite{Allc69}. However, backflow in quantum systems is too subtle and elusive to be directly detected experimentally~\cite{Year13,Brac94,Pacc20,Gous21,Barb23,Pate26}. Notably, it has been illustrated to be a universal interference effect linked to wave superoscillations~\cite{Berr06,Berr10}, which has made it accessible in optics~\cite{Elie20,Dani22,Ghos23,Xie25,Zhan25,He26}. 

Optical backflow falls into two categories: reversal of the canonical momentum (inverse local wavevector)~\cite{Elie20,Dani22,Ghos23,Zhan25,He26} and reversal of the Poynting momentum (energy backflow)~\cite{Novi07, Vave12, Mitr16, Khon18, Yuan19,Chen19,Kotl21,Xie25,Rado26}. These two distinct momenta respond differently to the electromagnetic field~\cite{Beli40,Berr09,Ohani86,BBN13,Anto16,Liu18}, so their reversal need not occur together. In addition, both categories include axial and angular forms. The axial forms have received particular attention~\cite{Novi07, Vave12, Mitr16, Khon18, Yuan19,Chen19,Kotl21,Xie25,Rado26}, whereas their angular counterparts remain less explored~\cite{Ghos24}; the angular (azimuthal) form is the subject of this work.

Azimuthal optical backflow was first reported by Ghosh et al.~\cite{Ghos23}, who superposed two scalar helical-phase Gaussian beams with different negative orbital angular momenta (OAM) and observed positive local OAM in the dark fringes. Subsequent studies extended this effect to a vectorial framework with two linearly polarized Laguerre-Gaussian beams~\cite{Xie25}, to single-photon weak measurement with two scalar Laguerre-Gaussian modes~\cite{Zhan25}, and to three scalar helical-phase Gaussian beams for higher-order symmetric patterns~\cite{He26}. Across these studies, the OAM involved is longitudinal, the light modes are scalar or linearly polarized, the azimuthal backflow is activated by tuning an amplitude ratio, and each examines only one of the two momentum densities. In particular, in the scalar description of Ref.~\cite{Ghos23, Zhan25}, equal amplitudes give no backflow and polarization is explicitly treated as inconsequential.

Transverse OAM (t-OAM), orthogonal to the propagation direction, has recently attracted attention~\cite{Blio12,Jhaj16,Hanc19,Chon20,Hanc21,Blio23,Porr23,Beks24,Vo26}. In our previous work, we constructed linearly polarized rotating-wave structured light carrying t-OAM and studied its spin-orbit coupling~\cite{Wang24}. A single such beam has a purely azimuthal, unidirectional circulation and therefore no azimuthal backflow. 

Two questions therefore arise: can both momentum densities be reversed in the azimuthal direction by the interference of t-OAM beams without amplitude tuning, and can this simultaneous reversal be gated by polarization alone? The present work answers both. Beyond the linearly polarized vectorial framework of Ref.~\cite{Xie25}, we endow each rotating-wave beam with an independently tunable polarization state and superpose two such beams with equal intensity and different topological charges. We show that the interference reverse both momentum densities in the azimuthal direction without amplitude tuning, and that this double reversal is switched by polarization alone. Under this configuration, both the azimuthal components of the Poynting and canonical momenta can exhibit backflow in the same angular sectors, but generally in different radial ranges.

This double azimuthal backflow is controlled solely by the polarization states of the beams: matched polarizations induce it, whereas mismatched ones suppress it. The double backflow is periodic in the azimuthal direction, with its angular periodicity fixed by the topological charge difference. These results establish polarization as a gate for azimuthal backflow, in which the polarization degree of freedom dictates the emergence of the double azimuthal backflow. Our findings may enable applications in polarization-switchable optical manipulation.

The paper is organized as follows. Section~\ref{II} recalls the single rotating-wave structured light and shows that its azimuthal momentum densities are unidirectional. Section~\ref{III} introduces the two-beam superposition, derives the polarization-gated double backflow, and establishes the angular locking to $\Delta\ell$. Section~\ref{IV} determines the radial intervals of the two reversals, and Sec.~\ref{IV} concludes with a discussion.

\section{Single rotating-wave structured light}\label{II}

We start from a rotating-wave structured light propagating along the azimuthal direction in the $(z,x)$ plane, which possesses a vortex of topological charge $\ell$ ($\ell=0,\pm1,\pm2,\dots$) and an intrinsic transverse $y$-directed OAM of $\ell\hbar$ per photon. The corresponding scalar rotating-wave wave function reads
\begin{equation}
\psi_{\ell}(\bm{r},t) =2\pi J_{\ell}(k\rho)e^{i\ell(\varphi+\frac{\pi}{2})-i\omega t},\label{eq:psi}
\end{equation}
in which $J_{\ell}(k\rho)$ denotes the $\ell$th-order Bessel function, $k=\omega/c$ is the wave number, with $\omega$ and $c$ being the frequency and the speed of light, respectively, and $(\rho,\varphi)$ are polar coordinates in the $(z,x)$ plane with $z=\rho\cos\varphi$ and $x=\rho\sin\varphi$. The sign of the topological charge $\ell$ fixes the azimuthal propagation direction, $+\varphi$ for $\ell>0$ and $-\varphi$ for $\ell<0$.

In our earlier work~\cite{Wang24}, two linearly polarized dual modes endowed with transverse OAM were built upon $\psi_\ell$: the TM mode, whose magnetic field points along $y$-direction,
\begin{equation}
\bm{B}_{\ell}^{\rm TM}=\psi_{\ell}\hat{\bm{y}},\qquad
\bm{E}_{\ell}^{\rm TM}=\frac{i}{k}\na\times\bm{B}_{\ell}^{\rm TM},
\label{eq:tm}
\end{equation}
together with its dual TE mode, whose electric field points along $y$-direction,
\begin{equation}
\bm{E}_{\ell}^{\rm TE}=-\psi_{\ell}\hat{\bm{y}},\qquad
\bm{B}_{\ell}^{\rm TE}=-\frac{i}{k}\na\times\bm{E}_{\ell}^{\rm TE},
\label{eq:te}
\end{equation}
where $\hat{\bm{y}}$ is the unit vector along $y$. Combining these two dual modes with a fixed $\pi/2$ phase difference and a tunable relative amplitude $\sigma$ gives
\begin{equation}
\{\bm{E}^{(\sigma,\ell)},\bm{B}^{(\sigma,\ell)}\}
=\frac{
\{\bm{E}_{\ell}^{\rm TM},\bm{B}_{\ell}^{\rm TM}\}
-i\sigma
\{\bm{E}_{\ell}^{\rm TE},\bm{B}_{\ell}^{\rm TE}\}
}{
\sqrt{1+\sigma^{2}}
}.\label{eq:spineb}
\end{equation}
The helicity index $\sigma\in[-1,1]$ governs the relative amplitude and handedness of the superposition: at $\sigma=0$ the linearly polarized TM mode of Ref.~\cite{Wang24} is retrieved, at $\sigma=\pm1$ the field becomes right- or left-handed circularly polarized, and for $0<|\sigma|<1$ it is elliptically polarized. Consequently, the transverse OAM of $\ell\hbar$ per photon is preserved throughout this polarization family, spanning linear, elliptical, and circular states~\cite{Wang26}.

The two momentum densities of interest are the Poynting momentum $\bm{\mathcal{P}}$ and canonical momentum $\bm{P}$:
\begin{equation}
\bm{\mathcal{P}}=gk\Re\big[\bm{E}^{*}\times\bm{B}\big],
\label{eq:poy}
\end{equation}
\begin{equation}
\bm{P}=\frac{g}{2}\Im\big(E_{j}^{*}\na E_{j}+B_{j}^{*}\na B_{j}\big).
\label{eq:can}
\end{equation}
Hereafter, Gaussian units are used with $g=(8\pi\omega)^{-1}$, and the index $j$ denotes a spatial index running over the three spatial coordinates. For later use, we quote the single-beam densities of the Poynting momentum, the canonical momentum, and energy, obtained by substituting Eqs.~\eqref{eq:spineb}--\eqref{eq:psi} into their definitions~\eqref{eq:poy}--\eqref{eq:can}:
\begin{align}
\bm{\mathcal{P}}^{(\sigma,\ell)}&=\frac{4\ell}{\rho}J_{\ell}^2\hat{\bm{\varphi}}+\frac{2k \sigma}{1+ \sigma^2}(J_{\ell-1}^2-J_{\ell+1}^2)\hat{\bm{y}},\label{eq:poyl}\\
\bm{P}^{(\sigma,\ell)}&=\frac{1}{\rho}\sum_{n=-1}^{1}2^{1-|n|}(\ell+n)J_{\ell+n}^2\hat{\bm{\varphi}}.\label{eq:pcl}
\end{align}
Hereafter, all momentum densities are expressed in units of the common factor $g\pi^2$. For brevity, the explicit $k\rho$ argument of the Bessel functions is omitted throughout; $\hat{\bm{\varphi}}$ is the unit vector in the azimuthal direction. As shown in the above equations, the canonical momentum density is $\sigma$-independent and points purely along the azimuthal direction in the $(z,x)$ plane. The Poynting momentum density, by contrast, develops an additional transverse $y$-directed component, which directly drives the emergence of the ``vortex-spin-momentum locking''~\cite{Wang26}; this transverse component does not enter the azimuthal backflow discussed here. Importantly, for such a single beam with fixed $\ell$, both azimuthal components $\mathcal{P}_{\varphi}^{(\sigma,\ell)}$ and $P_{\varphi}^{(\sigma,\ell)}$ are  $\sigma$- and ${\varphi}$-independent, are directed along the propagation direction and are unidirectional, neither reversing as $\varphi$ varies; no azimuthal backflow can arise from a single beam.

\section{Polarization-gated double azimuthal backflow and its angular locking}\label{III}

\begin{figure*}[htbp]
\centering
\includegraphics[width=1.95\columnwidth]{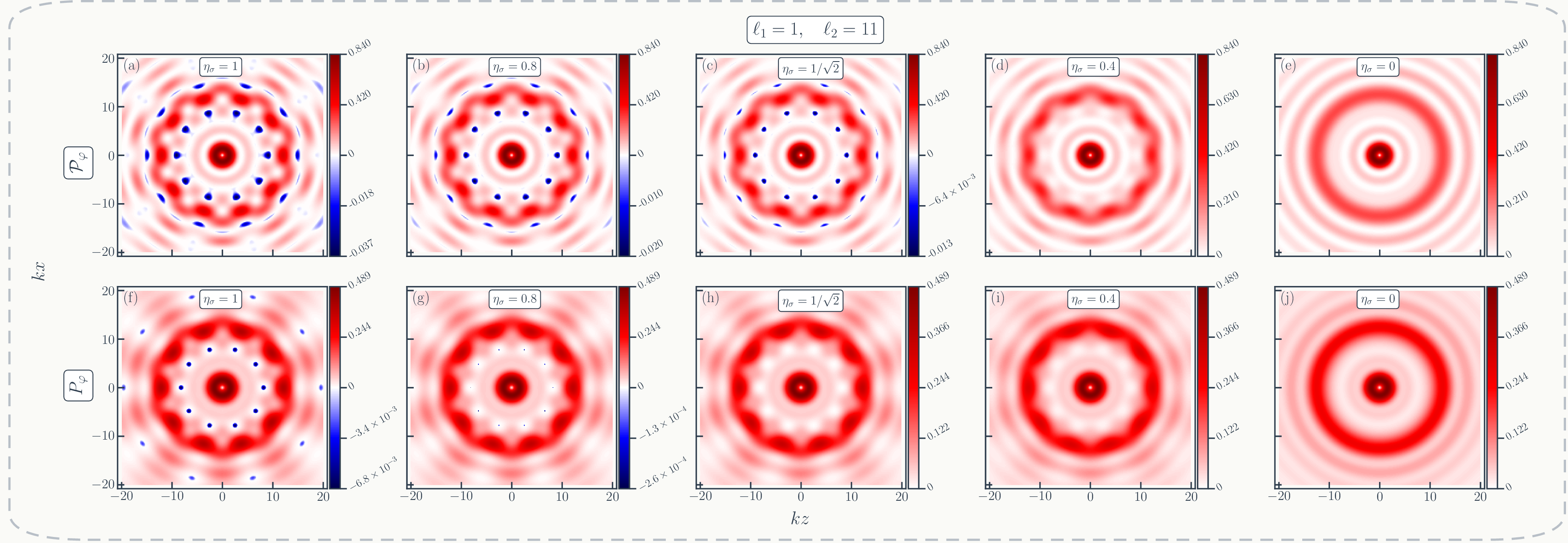}
\caption{Polarization-controlled gating of the double azimuthal backflow. Top row (a)--(e): azimuthal Poynting momentum $\mathcal{P}_{\varphi}$. Bottom row (f)--(j): azimuthal canonical momentum $P_{\varphi}$. Columns correspond to $\eta_\sigma=1$ ($\sigma_1=\sigma_2$; matched polarizations, gate open), $0.8$, $1/\sqrt{2}$ (e.g.\ one circular and one linear polarization), $0.4$, and $0$ ($\sigma_1=-\sigma_2=\pm1$; opposite-handed circular polarizations, gate closed). At $\eta_\sigma=1$ both momentum densities reverse sign in the same angular sectors; as $\eta_\sigma$ decreases the cross terms are attenuated but finite, and the backflow weakens; at $\eta_\sigma=0$ the cross terms vanish, both momentum densities reduce to the sum of two single-beam circulations, and the backflow is gated off. All plots are for equal-intensity beams with $\ell_1=1$, $\ell_2=11$.}
\label{fig:match}
\end{figure*}

We now consider the superposition of two t-OAM beams of equal total intensity whose topological charges and polarization states are both different. We fix equal total intensities for the two beams, so that all gating is carried by polarization alone and no amplitude tuning enters; the common normalization factor is omitted in what follows. Let the two beams have topological charges $\ell_{1}$ and $\ell_{2}$ and helicity indices $\sigma_{1}$ and $\sigma_{2}$. Since both momentum densities of a single beam are strictly unidirectional in the azimuthal direction (Sec.~\ref{II}), a nonzero interference-induced backflow requires the two beams to circulate in the same sense, i.e.\ $\ell_1\ell_2>0$; for $\ell_1\ell_2<0$ the single-beam terms already have opposite signs and no interference-induced reversal occurs~\cite{Ghos24}. Without loss of generality we take $\ell_2>\ell_1>0$, the negative-charge case following by reflection, $\ell\to-\ell$ being equivalent to $\varphi\to-\varphi$. The resulting fields take the form
\begin{equation}
\bm{E}=\bm{E}^{(\sigma_{1},\ell_{1})}+\bm{E}^{(\sigma_{2},\ell_{2})},
\quad
\bm{B}=\bm{B}^{(\sigma_{1},\ell_{1})}+\bm{B}^{(\sigma_{2},\ell_{2})}.
\label{eq:super}
\end{equation}
Substituting Eq.~\eqref{eq:super} into Eqs.~\eqref{eq:poy} and~\eqref{eq:can} separates each azimuthal momentum density into single-beam and interference contributions:
\begin{align}
\mathcal{P}_{\varphi}&=\mathcal{P}_{\varphi}^{(\sigma_1,\ell_1)}+\mathcal{P}_{\varphi}^{(\sigma_2,\ell_2)}+\mathcal{P}_{\varphi}^{\times},\label{Ptot}\\
P_{\varphi}&=P_{\varphi}^{(\sigma_1,\ell_1)}+P_{\varphi}^{(\sigma_2,\ell_2)}+P_{\varphi}^{\times},
\end{align}
with
\begin{equation}
 \mathcal{P}_{\varphi}^{\times}=4\eta_{\sigma}\frac{\ell_1+\ell_2}{\rho}J_{\ell_1}J_{\ell_2}\cos\alpha,\label{Pcross}  
\end{equation}
\begin{equation}
  P_{\varphi}^{\times}=\frac{\eta_{\sigma}}{\rho}\sum_{n=-1}^{1}2^{1-|n|}(\ell_1+\ell_2+2n)J_{\ell_1+n}J_{\ell_2+n}\cos\alpha,\label{Pccross}  
\end{equation}
\begin{equation}
\eta_{\sigma}=\frac{1+\sigma_1\sigma_2}{\sqrt{(1+\sigma_1^{2})(1+\sigma_2^{2})}},\label{eq:overlap}
\end{equation}
where $\alpha=\Delta\ell\,(\varphi+\pi/2)$ and $\Delta\ell=\ell_2-\ell_1>0$. Equation~\eqref{eq:overlap} measures the overlap of the two polarization states within the TM--TE basis of Eq.~\eqref{eq:spineb}. It gives $\eta_\sigma=1$ when $\sigma_1=\sigma_2$ and vanishes only at $\sigma_1=-\sigma_2=\pm1$; intermediate values correspond to partial overlap.

A key feature of Eqs.~\eqref{Pcross} and~\eqref{Pccross} is that both cross terms share the same two factors. First, both are proportional to the same polarization overlap $\eta_\sigma$, which is absent from the single-beam terms in Eq.~\eqref{Ptot}; the polarization therefore acts as a common gate on both interference contributions. Second, both carry the same angular factor $\cos\alpha$, so two backflow sectors have the same angular structure and share the same angular period and the same angular centers. The shared factors thus fix a common angular structure, while the radial dependence, carried by distinct Bessel combinations, remains different for the two momenta.

\textbf{Polarization gating.} Since $\eta_\sigma$ is contained in both cross terms and is absent from the single-beam terms, it acts as a continuous gate on the interference. At $\eta_\sigma=1$, i.e.\ $\sigma_1=\sigma_2$ (matched circular, linear, or elliptical polarizations), the cross terms are maximal and both $\mathcal{P}_{\varphi}$ and $P_{\varphi}$ reverse sign in the same angular centers, so the double backflow is fully developed [Figs.~\ref{fig:match}(a),(f)]. As $\eta_\sigma$ decreases from unity, the cross terms are attenuated but remain finite, and the backflow weakens continuously for each density [Fig.~\ref{fig:match}(b)--(d),(g)--(i)]. At $\eta_\sigma=0$, i.e.\ $\sigma_1=-\sigma_2=\pm1$ (two circularly polarized beams of opposite handedness), the cross terms vanish identically, both momentum densities reduce to the sum of two unidirectional single-beam circulations, and no backflow occurs [Fig.~\ref{fig:match}(e),(j)]. The two endpoints $\eta_\sigma=1$ and $0$ thus form a binary gate, with the intermediate cases showing a continuous polarization-controlled weakening.

\begin{figure*}[htbp]
\centering
\includegraphics[width=1.95\columnwidth]{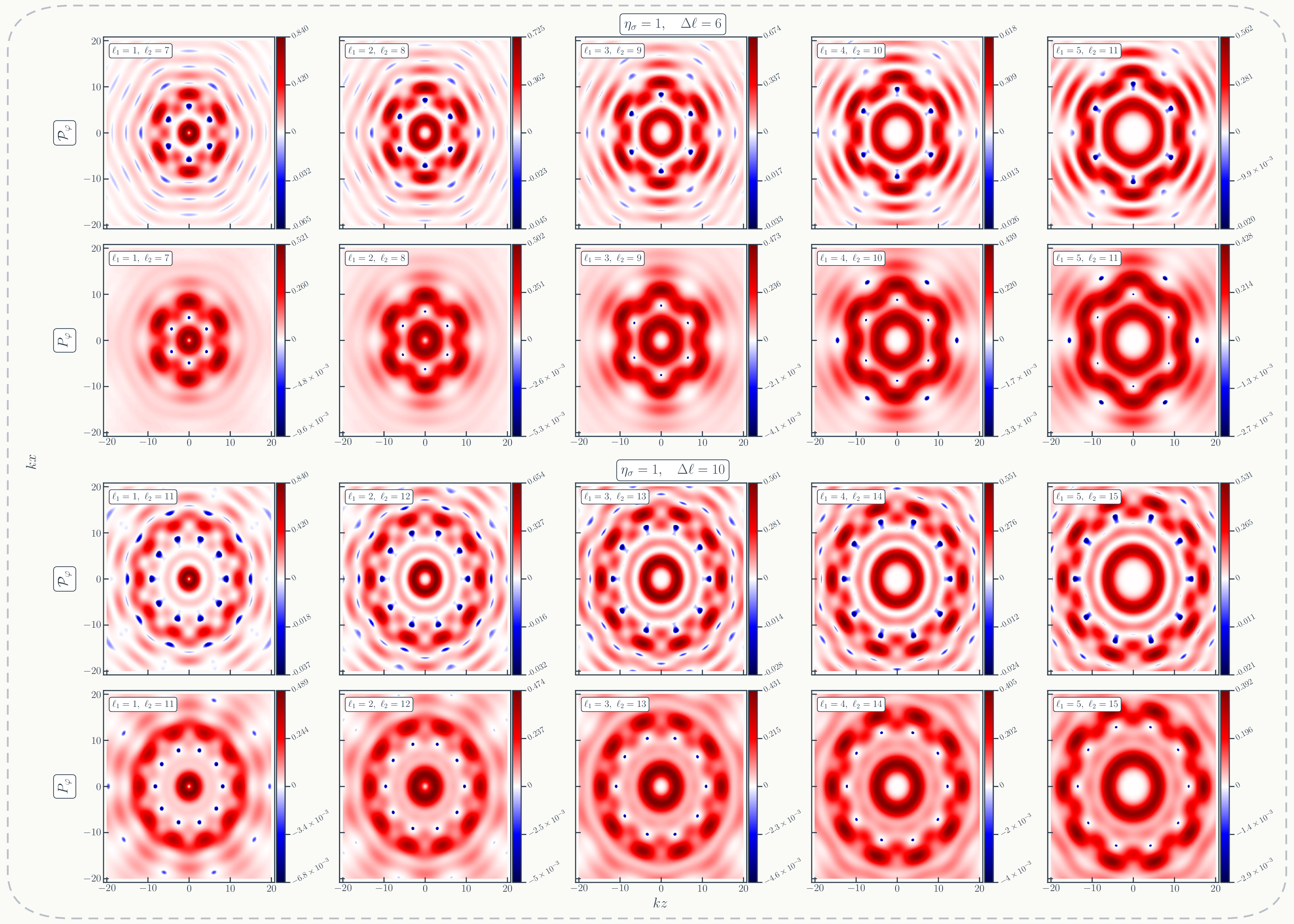}
\caption{Angular periodicity of the backflow locked to the charge difference $\Delta\ell$. Rows 1 and 2: $\mathcal{P}_{\varphi}$ and $P_{\varphi}$ for $\Delta\ell=6$; rows 3 and 4: $\mathcal{P}_{\varphi}$ and $P_{\varphi}$ for $\Delta\ell=10$. Columns 1--5: $(\ell_1,\ell_2)=(1,7),(2,8),(3,9),(4,10),(5,11)$ (rows 1--2) and $(1,11),(2,12),(3,13),(4,14),(5,15)$ (rows 3--4), all with matched polarizations ($\eta_\sigma=1$).}
\label{fig:delta}
\end{figure*}

\textbf{Angular locking.} Since the cross terms depend on $\varphi$ only through $\cos(\Delta\ell\,\varphi+\Delta\ell\,\pi/2)$, the backflow regions are periodic in $\varphi$ with period $\Lambda_\varphi=2\pi/\Delta\ell$, fixed by $\Delta\ell$ alone and independent of the individual charges $\ell_1$ and $\ell_2$. Their angular extrema lie at $\cos\alpha=\pm1$, which determines the angular centers of backflow regions: a series of backflow centers sits at $\cos\alpha=+1$ or $\cos\alpha=-1$, with each backflow center at $\cos\alpha=+1$ separated from the neighboring one at $\cos\alpha=-1$ by $\pi/\Delta\ell$. Since $\eta_\sigma$ does not enter the phase $\alpha=\Delta\ell(\varphi+\pi/2)$, changing it at fixed $(\ell_1,\ell_2)$ leaves the angular positions and the periodicity of the backflow regions unchanged; it only rescales the two cross terms and hence the area of the backflow regions. Figure~\ref{fig:delta} confirms the periodicity: for matched polarizations, the number of backflow regions in the azimuthal direction increases with $\Delta\ell$ and their angular period is $2\pi/\Delta\ell$, independent of $\ell_1$ and $\ell_2$.

Two parameters therefore control the effect: the polarization overlap $\eta_{\sigma}$ determines whether and how strongly the double backflow appears, while the charge difference $\Delta\ell$ determines where it appears and with what period in the azimuthal direction. We stress that the two reversals share the same angular sector but not necessarily the same radial interval: at intermediate $\eta_\sigma$ one momentum density may cease to backflow while the other still does (see Sec.~\ref{IV}).

\section{Radial distribution of backflow regions}\label{IV}

\begin{figure}[htbp]
\centering
\includegraphics[width=0.85\columnwidth]{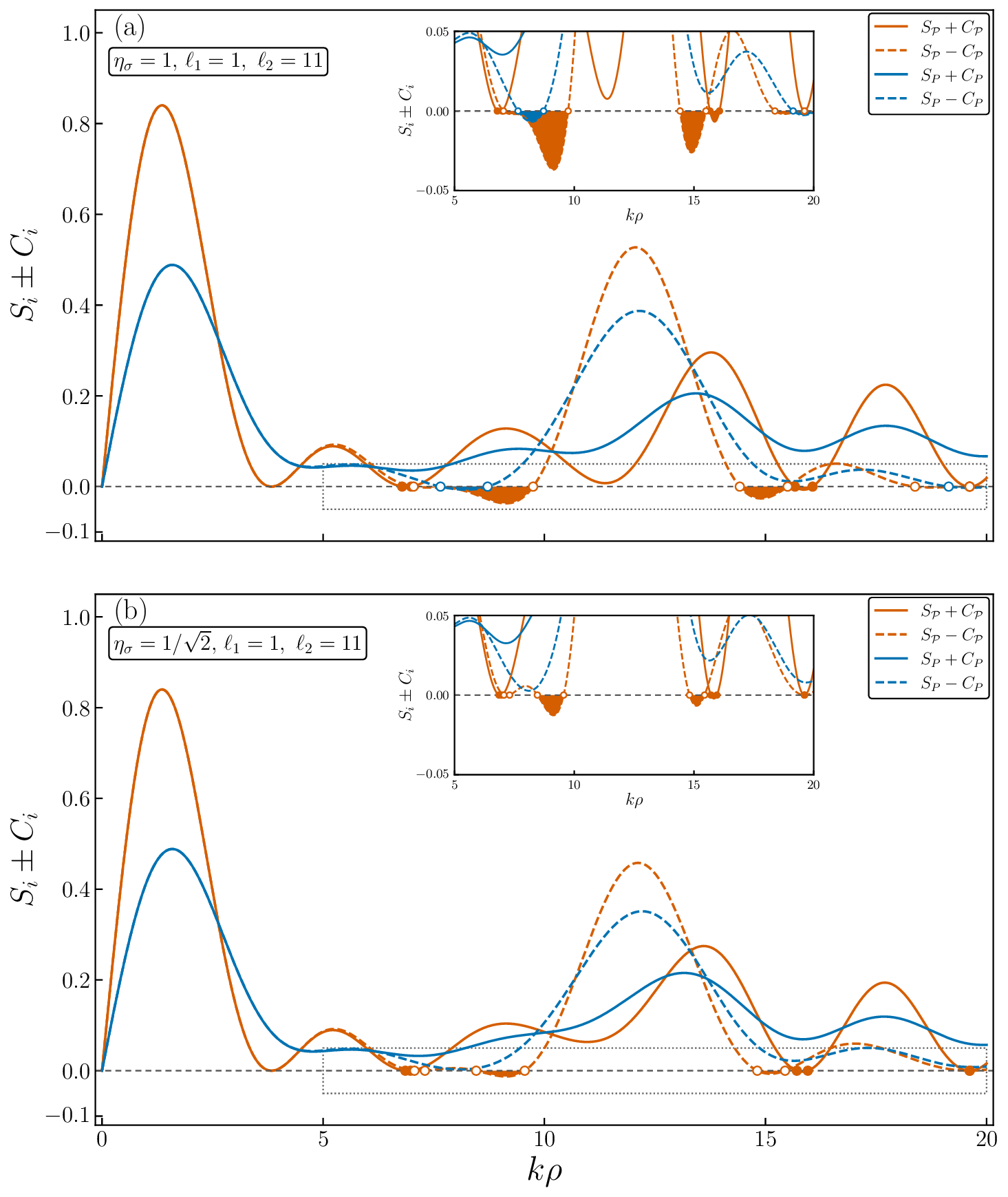}
\caption{Radial distribution of backflow intervals with $\ell_1=1$, $\ell_2=11$. (a) Matched polarizations ($\eta_\sigma=1$); (b) Partially matched polarizations ($\eta_\sigma=1/\sqrt{2}$). Plotted versus normalized radial coordinate $k\rho$ are $S_{\mathcal{P}}\pm C_{\mathcal{P}}$ (red) and $S_{P}\pm C_{P}$ (blue); solid and dashed curves denote the $+$ and $-$ branches, respectively. Backflow occurs in radial intervals satisfying $|C_i|>S_i$ (Eq.~\ref{eq:crit}), with intervals shaded red for $\mathcal{P}_\varphi$ and blue for $P_\varphi$. Filled and open circles mark the zeros of $S_i+C_i$ and $S_i-C_i$, respectively (red for $\mathcal{P}_\varphi$, blue for $P_\varphi$); these zeros correspond to the boundaries at $\cos\alpha=-1$ and $\cos\alpha=+1$ and trace out the edges of the backflow intervals. The inset provides a magnified view of the $k\rho\in[5,20]$ range (dotted rectangle in the main panel) to resolve the small backflow lobes.}
\label{fig:radial}
\end{figure}

Having localized the azimuthal positions of backflow in Section~\ref{III}, we now turn to determining the radial ranges over which each momentum density is reversed. Equations~\eqref{Pcross} and~\eqref{Pccross} can be written as $\mathcal{P}_{\varphi}^{\times}=C_{\mathcal{P}}(\rho)\cos\alpha$ and $P_{\varphi}^{\times}=C_{P}(\rho)\cos\alpha$, with the radial coefficients
\begin{equation}
C_{\mathcal{P}}(\rho)=4\eta_\sigma\frac{\ell_1+\ell_2}{\rho}J_{\ell_1}J_{\ell_2},
\end{equation}
\begin{equation}
C_{P}(\rho)=\frac{\eta_\sigma}{\rho}\sum_{n=-1}^{1}2^{1-|n|}(\ell_1+\ell_2+2n)J_{\ell_1+n}J_{\ell_2+n},
\end{equation}
and the sums of the two single-beam terms
\begin{align}
S_{\mathcal{P}}(\rho)& = \mathcal{P}_\varphi^{(\sigma_1,\ell_1)}+\mathcal{P}_\varphi^{(\sigma_2,\ell_2)},\\
S_{P}(\rho)& = P_\varphi^{(\sigma_1,\ell_1)}+P_\varphi^{(\sigma_2,\ell_2)}.
\end{align}
Equations~\eqref{eq:poyl} and~\eqref{eq:pcl} show that $S_i(\rho)>0$ for $\ell_2>\ell_1\ge1$, while $S_i$ and $C_i$ both vary with $\rho$.

\textbf{Backflow condition.} At fixed $\rho$, the azimuthal density of species $i=\mathcal{P},P$ varies linearly with $\cos\alpha\in[-1,1]$ and takes values in $[S_i-|C_i|,\,S_i+|C_i|]$. Backflow therefore occurs at this $\rho$ if and only if
\begin{equation}
|C_i(\rho)|>S_i(\rho),
\label{eq:crit}
\end{equation}
in which case it occupies the angular range
\begin{equation}
\operatorname{sgn}(C_i)\,\cos\alpha<-\frac{S_i(\rho)}{|C_i(\rho)|},
\end{equation}
i.e.\ a sector centered on $\cos\alpha=-\operatorname{sgn}(C_i)$.

\textbf{Radial extent.} The radial boundaries are set by $|C_i(\rho)|=S_i(\rho)$, not by the zeros of $C_i(\rho)$; at a zero of $C_i$ the full density equals $S_i>0$ and no backflow occurs. Since $S_i$ and $C_i$ vary differently with $\rho$ for the two species, the radial intervals satisfying Eq.~\eqref{eq:crit} generally differ between $\mathcal{P}_\varphi$ and $P_\varphi$: the double backflow occupies a common angular sector but generally different radial ranges. For $\ell_1=1$, $\ell_2=11$, at the matched endpoint $\eta_\sigma=1$ which switches the double backflow on, Fig.~\ref{fig:radial}(a) shows the radial intervals of double backflow shaded red for $\mathcal{P}_\varphi$ and blue for $P_\varphi$ (the inset magnifies the small backflow lobes in the radial direction).

Because $S_i$ is independent of $\eta_\sigma$ while $C_i\propto\eta_\sigma$, reducing $\eta_\sigma$ from unity rescales only the left-hand side of Eq.~\eqref{eq:crit} and leaves the single-beam sums untouched; the radial boundaries are thus set by the Bessel structure at $\eta_\sigma=1$ and move inward as $\eta_\sigma$ decreases. Keeping the same case of $\ell_1=1$ and $\ell_2=11$, Fig.~\ref{fig:radial}(b) shows the partially matched polarization case ($\eta_\sigma=1/\sqrt{2}$): the reduction is sufficient to suppress backflow for the canonical momentum density, with $|C_P|$ no longer exceeding $S_P$ at any $\rho$, so that all backflow intervals of $P_\varphi$ disappear. The Poynting momentum density $\mathcal{P}_\varphi$, by contrast, still exhibits backflow in this example, since its larger cross-term coefficient $C_{\mathcal{P}}$ remains sufficient to satisfy $|C_{\mathcal{P}}|>S_{\mathcal{P}}$ over a finite radial range.  This is consistent with the sequence in Figs.~\ref{fig:match}(c) and (h).

\section{Discussion and Conclusion}
In summary, we have shown that superposing two t-OAM beams with distinct transverse topological charges produces azimuthal backflow in their interference, and that this backflow appears simultaneously in the Poynting and canonical momentum densities. The effect is controlled by two parameters: the polarization overlap $\eta_\sigma$, set by the helicity indices $\sigma_1,\sigma_2$, acts as a common gate on both interference cross terms and hence on the existence of the double backflow in a common angular sector; the topological charge difference $\Delta\ell$ fixes the angular periodicity of the backflow regions. The radial intervals of the two reversals are generally different and may even separate at intermediate $\eta_\sigma$.

Unlike previous schemes, in which backflow was activated by an amplitude ratio and the reversals of the Poynting and canonical momentum densities were not tied to a common parameter~\cite{Ghos23,Xie25,Zhan25,He26}, here the two beams have equal intensities and the two reversals share a single origin in the polarization overlap of Eq.~\eqref{eq:overlap}. This makes the double backflow gated on and off by the polarization of the two beams, without altering the topological charges or the amplitudes.

The double backflow is exact for matched polarizations ($\sigma_1=\sigma_2$, circular, linear, or elliptical), where both momentum densities inherit the same $\Delta\ell\,\varphi$ modulation. For intermediate or mismatched polarizations the interference terms are reduced or suppressed, and the simultaneous reversal weakens continuously: the two momentum densities remain individually well defined, but their common reversal is lost. This continuous weakening is itself potentially useful, since it allows the double backflow to be isolated or suppressed without altering the topological charges or the amplitudes. The polarization-switchable double backflow thus offers a route to control azimuthal backflow in optical manipulation, where the mechanical response of a particle could be switched on and off by changing the polarization of the two beams. 

\section*{Acknowledgments}
The author sincerely acknowledge Prof. Chen for his invaluable guidance and support during this work.

\end{document}